# Growing green: the role of path dependency and structural jumps in the green economy expansion

Seyyedmilad Talebzadehhosseini[a,c], Steven R. Scheinert[a,c], and Ivan Garibay[a,b,c]

[a] *Department of Industrial Engineering and Management Systems, University of Central Florida, Orlando, Florida, United States of America*

[b] *Computer Science Department, University of Central Florida, Orlando, Florida, United States of America*

[c] *Complex Adaptive Systems Laboratory, University of Central Florida, Orlando, Florida, United States of America*

**Corresponding author**:
Dr. Ivan Garibay
Assistant Professor
Department of Industrial Engineering and Management Systems
College of Engineering and Computer Science
University of Central Florida
4000 Central Florida Blvd., Orlando, FL 32816-2993 USA
Room 424 Engineering II (Building 91)
Telephone: (407) 882-1163
Email: igaribay@ucf.edu

**ABSTRACT**

Existing research argues that countries increase their production basket by adding products which require similar capabilities to those they already produce, a process referred to as path dependency. Green economic growth is a global movement that seeks to achieve economic expansion while at the same time mitigating environmental risks. We postulate that countries engaging in green economic growth are motivated to invest strategically to develop new capabilities that will help them transition to a green economy. As a result, they could potentially increase their production baskets not only by a path dependent process but also by the non-path dependent process we term "high-investment structural jumps". The main objective of this research is to determine whether countries increase their green production basket mainly by a process of path dependency, or alternatively, by a process of structural jumps. We analyze data from 65 countries and over a period from years 2007 to 2017. We focus on China as our main case study. The results of this research show that countries not only increase their green production baskets based on their available capabilities, following path dependency, but also expand to products that path dependency does not predict by investing in innovating and developing new environmental-related technologies.

**Keywords**: product space, path-dependence, non-path dependence, green economy

**Introduction**

Countries are faced with the paired challenges of both growing their economics and mitigating environmental risks (Knight and Schor, 2014). In most public debates, these two goals are posed as being in competition with each other. A growing call for green economic development, including recent efforts towards the Green New Deal proposed in the United States Congress, show how countries can pursue both goals simultaneously and in a mutually supportive manner (Andersen, 2018; He et al., 2019; Fraccascia et al., 2018; Han, 2018). The expansion of a country's green production basket enables it to grow and develop its economy while meeting its environmental needs (McAfee, 2016; D'Amato et al., 2017; Weber and Cabras, 2017). According to the path dependency hypothesis (Hidalgo et al., 2007), an economy that is not prepared to make this expansion will require a more difficult transition to expand its capacity. Examining growth patterns in those countries that are building and expanding a green economic sector will show whether expansion into and within the green economy adheres to the path dependency hypothesis.

Path dependency is driven by product similarities. Hausmann and Klinger (2007) and Hidalgo et al. (2007) established the concept of a Product Space (PS) that provides a measure of similarity between any two products in an economy. This measure is based on the idea that if two products share similar production inputs, such as resources, technology, or infrastructure, then they have a greater similarity to products that do not share any of these inputs. Hidalgo et al (2007) showed that it is also more likely that an economy will produce both products and that expansion is more likely to add products similar to what the economy already produces, establishing the path dependency hypothesis. Mealy and Teytelboym (2018) proposed the concept of a Green Product Space (GPS) by constructing a comprehensive list of green products and linking all green products as a network by defining a measure of green products' similarities based on Hausmann and Klinger (2007) and Hidalgo et al. (2007)'s definitions. The GPS allows the application of concepts that Hidalgo et al. (2007) examined in the full PS to a country's green economy.

Diversifying production according to path dependency tends to generate faster economic growth (Hausmann and Klinger, 2007; Hidalgo et al., 2007; Mealy and Teytelboym, 2018; Coniglio et al., 2018). Path-defying diversification is rare, especially for poorer countries (Hidalgo et al., 2007; Coniglio et al., 2018). Though it is harder and riskier for the economy to jump in their PS and GPS networks, the required structural changes can be expected to generate greater economic development over a longer period of time than path dependent diversification despite its faster GDP growth. (Coniglio et al., 2018) have demonstrated that product diversification across all goods tends to follow path dependency. However, this has not been examined for the expansion of green production and the diversification of an economy's basket of green products.

This research asks if empirically observed patterns of green product diversification follow the path dependency hypothesis as suggested by applying Hidalgo et al.'s (2007) work in the PS framework to the GPS framework (Coniglio et al.2018; Mealy and Teytelboym, 2018). This analysis allows us to answer the following question:

*Does a country's ability to diversify its green product basket follow a path-dependent or non-path dependent process?*

The answer to the above question will help us understand the reasons why some countries, such as the United States, China, and Germany, had successful transitions to a green economy (and a more diversified green production basket) and thus stand as the top leaders in green economy growth. The goals of this research are to identify the patterns of green growth in all countries with a special focus on China (the country with the highest green product export values according to Mealy and Teytelboym [2018]), and the factors that enable these countries to have structural jumps in their green production baskets and stand as leaders in developing their green economies. The results of this research can be used as a suggestion for a possible direction that countries with a lower rate of green growth can take to diversify their green production basket.

**Literature**

Economic growth is not a sufficient indicator of the level of economic development in a country (Moyo, 2009; Acemoglu et al. 2002). Economic growth is simply the value of the annual increase in a country's production basket along with the growth rate of its Gross Domestic Product (GDP) (Zhang and Zeng, 2008). Economic development does not only happen through changing existing production materials, but through several changes in a country's production structure (Schumpeter, 1961). Structural changes in a country's production materials enables it to enhance its production basket and achieve its desired economic development (Yang, 1990). This development causes countries to face several problems, such as environmental risks (Gu et al., 2018; Lederer et al., 2018; Knight and Schor, 2014), and high consumption of energy and natural resources (OECD, 2013).

According to Knight and Schor (2014), climate change is the most serious environmental issue that the world is facing today. Studies show the strong positive relation between a country's economic development and its carbon dioxide emissions (Knights and Schor, 2014; Jorgenson and Clark; 2012; York et al., 2003; Dietz and Rosa; 1997). Therefore, it is important to limit global climate change "…below the critical threshold of 2 C" (Lorek and Spangenberg, 2014, p. 2). Green growth and the transition to green economy enables a country to reduce its environmental risks, and address some of its economic, social, and environmental challenges (Hoffmann, 2011).

**Path Dependency in Product Diversification**

The concept of path dependency in product diversification was proposed by Hidalgo et al. (2007). They argued that a country's economy grows by producing more products, and it is easy for a country to produce products that require similar capabilities, such as capital, technology, infrastructure, and labor, of the country's existing export basket. According to Coniglio et al.

(2018), "current production capabilities are the key link between what a country produces today and what it will produce tomorrow, in other words the essence of the mechanism of path-dependence…" (p. 10). Hausmann and Klinger (2010) and Hidalgo (2012) analyzed the export baskets of Ecuador and some African countries— Kenya, Mozambique, Rwanda, Tanzania and Zambia— and showed that they inhabit a peripheral position in the PS. Coniglio et al (2018) added that this peripheral position is persistent over time. Coniglio et al. (2018) discussed that, Minondo (2011) studied export baskets of 91 countries to show how they are diversified in their production baskets by calculating the degree of centrality in the PS. The results of their research showed that the degree of centrality in the PS "…is a strong predictor of diversification level" (p. 10). Coniglio et al. (2018) mentioned another study that investigates the path-dependency in countries' products diversification and is proposed by Boschma and Capone (2016) where they "…analyzed the process of trade diversification for EU-27 and European Neighbourhood Policy (ENP) countries between 1995 and 2010. The authors find evidence of path-dependence as countries develop their revealed comparative advantage in products related to those in which they were already specialized" (p. 10).

Mealy and Teytelboym (2018) showed that a country's green growth and green development also follow a path-dependence process. The authors showed that countries can expand their green production baskets based on their existing green production capabilities and used this finding as a fundamental basis regarding how countries can re-orient their current industrial capabilities in order to have a better green growth. Their work remains one of only a few works that examine the dynamics described above in the GPS.

**Green Economy and Green Growth**

According to Knight and Schor (2014) and the United Nations Environment Programme (UNEP), a green economy can be defined as "one that results in improved human well-being and social equality, while significantly reducing environmental risks and ecological scarcities" (p. 3723). On the other hand, green growth is defined as "growth achieved by saving and using energy and resources efficiently to reduce climate change and damage to the environment, securing new growth engines through research and development of green technology, creating new job opportunities, and achieving harmony between the economy and environment" (Kasztelan, 2017, p. 489). Green growth and green economy are both suggested as solutions to financial and economic crises (Kasztelan, 2017; Lane, 2010). Both involve improving the global economy by investing in the environmentally friendly products, markets, and services (Kasztelan, 2017; Lane, 2010). Although the terms green economy and green growth have different origins, they are often used interchangeably (Kasztelan, 2017).

The driving force behind the development of green economy and green growth is their high focus on comprehensively incorporating the environment in the economy. Mainly through technological innovations, in the concepts of green economy and green growth, feasible approaches to improve the results of economic activity are identified while considering climatic problems and deficiency in natural resources (Kasztelan, 2017). Under the green economy approach, the goal is two-fold. It aims at transforming the economy in such a way that it reduces environmental and ecological deficiencies, while at the same time improving justice and social welfare. Such change will be achieved by investment, the creation of "green" jobs, the creation of markets for new products, and the reinforcement of international trade. The main goal under green growth is to maintain the economic growth, while taking into account the importance of natural capital and recognizing its role in production (Kasztelan, 2017).

**Product Space**

The concept of PS is defined by Hausmann and Klinger (2007) and Hidalgo et al. (2007). The PS is a network between 774 products that are produced and traded between all countries in the world. Products are related when they use similar inputs of capital, recourses, and labor, and the PS network connects these products according to their relatedness. The relatedness between any two products $i$ and $j$ in the PS is calculated as "…the minimum of the pairwise conditional probability of being co-exported…" (Coniglio et al., 2018, p. 10) with revealed comparative advantage (RCA ≥ 1):

$$\varphi_{i,j} = \min\{P(x_i \ / \ x_j), P(x_j \ / \ x_i)\}$$

where $\varphi_{i,j}$ is the measure of relatedness between any two products $i$ and $j$, $x_i$ and $x_j$ are the export values of products $i$ and $j$ that have been traded between countries, respectively, and $P(x_i \ / \ x_j), P(x_j \ / \ x_i)$ "…is the conditional probability of exporting good $i$ given that you export good $j$" (Hidalgo et al., 2007, p. 2). According to Hidalgo et al. (2007), the new product $j$ that will be added to a country's export basket is the one that has the highest relatedness with the products that are already produced and exported. The main hypothesis of the PS is that "the evolution of countries comparative advantage can be represented over the PS as gradual 'jumps' from one node that represents a product already in the export basket to the closest nodes not in the production basket, that is products in which countries have a latent comparative advantage from sharing the use of similar production capabilities" (Coniglio et al., 2018, p. 6). More explicitly, the path dependency hypothesis claims that countries tend to grow their economy by adding products that are similar to those for which the country already produces with high RCA.

**Green Product Space**

Mealy and Teytelboym (2018) used the concept of PS and developed a network of relatedness between 293 green products—called the GPS— that are traded between 1995 and 2014. The main hypothesis of GPS is that countries tend to develop their green economy according to their existing green production capabilities. Similar to Hidalgo et al. (2007), Mealy and Teytelboym (2018) argued that the next new green product that is to be added to a country's green production basket is the one that has the highest relatedness value with the green products that are already produced and exported in the green production baskets of that country. In addition, Mealy and Teytelboym (2018) ranked countries according to their Green Complexity Index (GCI) and showed countries with high GCI, "…have higher environmental patenting rates, lower $CO_2$ emissions, and more stringent environmental policies" (Mealy and Teytelboym, 2018, p. 1).

They further constructed the Green Adjacent Possible (GAP) measure that "…represents the set of technologically proximate green products that a country could potentially become competitive in" (Mealy and Teytelboym, 2018, p. 1). Finally, the authors constructed a measure— Green Complexity Potential (GCP)— to predict "…countries' future competitiveness in green products" (Mealy and Teytelboym, 2018, p. 1), and show that the relation between GCP and GCI will "…[suggest] the path-dependence in the accumulation of green production capabilities" (Mealy and Teytelboym, 2018, p. 1).

**Revealed Comparative Advantage (RCA)**

The Revealed Comparative Advantage (RCA) index can be defined as "a measure of the relative ability of a country to produce a good vis-`a-vis its trading partners" (French, 2017, p. 83) and is defined by Balassa (1965). Hausmann et al. (2014) described Balassa (1965)'s definition of RCA as by saying that a country has RCA on a product if it produces and exports the product more than "…a fair share, that is, a share that is equal to the share of total world trade that the product [represents]" (Hausmann et al., 2014, p. 25). The RCA of country c, for product i, can be calculated as (Hausmann et al., 2014; Fraccascia et al., 2018):

$$RCA_{ci} = \frac{X_{ci}}{\sum_j X_{cj}} / \frac{\sum_c X_{ci}}{\sum_{ci} X_{cj}} \geq 1$$

where $X_{ci}$ is the export value of product *i* for country *c*, $\sum_j X_{cj}$ is the total export value of all products, *j*, that is exported by country, *c*, $\sum_c X_{ci}$ is the total export value of product *i* that is exported by all countries *c*, and $\sum_{ci} X_{cj}$ is the total value of all products that has been traded between all countries in the world. (Hausmann et al., 2014) then used this measure to develop "…a matrix that connects each country to the products that it makes" (Hausmann et al., 2014, p. 25). The matrix values can be calculated as (Hausmann et al., 2014):

$$M_{ci} = \begin{Bmatrix} 1 & if\ RCA_{ci} \geq 1; \\ 0 & otherwise. \end{Bmatrix}$$

where $M_{ci}$ is the entries in the matrix and it is 1 if country c exports product i, with Revealed Comparative Advantage larger than 1, and 0 otherwise.

**Data**

To obtain the data for this research, first a comprehensive list of green products must be developed. To this end, the list of green products is obtained from the Organization for Economic Cooperation and Development (OECD), and with this, a comprehensive list of 247 green products is constructed (Sauvage, 2014)[1]. The green products in this comprehensive list are classified based on the 6-digit Harmonized System (HS). Second, the green growth indicators and their related data are collected from the OCED statistics database to explore whether these indicators play a role on enhancing the countries' ability to have structural jumps in their green product space or not. The green growth indicators can be listed as: development of environmental-related technologies per capita, number of patents related to environmental-related technologies, and development of environment-related technologies, as a % of all technologies. In the third step, the trade data for 247 green products were obtained from the United Nation Comtrade database (UN Comtrade, 2019). The data includes:

1. Information on the year that each green product was exported
2. The countries that exported and imported green products
3. All green product codes according to HS classification.
4. Each country's code
5. Trade values of each green product for all countries that was traded in each year (2007 to 2017).

---

[1] List of green products can be found at https://www.oecd-ilibrary.org/trade/the-stringency-of-environmental-regulations-and-trade-in-environmental-goods_5jxrjn7xsnmq-en

The trade value of each green product shows how much a specific green product was exported by each country. The trade value is based on US Dollars, and for all 247 green products the export values are considered. Trade values are used to calculate the RCA and the relatedness for each country per year. A list of some green products and their related data is listed in Table 1.

Table 1. List of some green products

| Year | Country | Country ISO code | HS6 green products code | Green product name | Export value (US$) |
|---|---|---|---|---|---|
| 2013 | Germany | DEU | 390940 | Phenolic resins, in primary forms | $251,471,129 |
| 2013 | United States of America | USA | 390940 | Phenolic resins, in primary forms | $237,674,965 |
| 2013 | China | CHN | 390940 | Phenolic resins, in primary forms | $156,046,507 |
| 2017 | Germany | DEU | 840690 | Turbines; parts of steam and other vapor turbines | $500,354,255 |
| 2017 | United States of America | USA | 840690 | Turbines; parts of steam and other vapor turbines | $309,129,267 |
| 2017 | China | CHN | 840690 | Turbines; parts of steam and other vapor turbines | $430,767,939 |

**Methods**

This analysis uses seven steps to answer the research question listed above:

1. In order to begin the analysis, two years should be defined to understand the country's pattern of green growth. In this research, the initial year is 2007, $t_0=2007$ and final year is 2017, $t_1=2017$.

2. Define the new green products.

3. Calculate the relatedness between any pair of products *i* and *j* that was traded in $t_0=2007$ to construct an MxM matrix of relatedness between all products *i* and *j*.

4. Construct the set of products with RCA>1 at $t_0=2007$ for each country.

5. Calculate the relatedness between new green products and all products with RCA>1 at $t_0=2007$.

6. Construct the matrix of relatedness, M$x$G, between new green products and all products with RCA>1 at $t_0=2007$.

7. Provide statistical analysis on the obtained data at step 6 to understand whether or not countries followed path-dependence process to grow their green economy.

We base our approach in the PS framework (Hidalgo et al., 2007) and dart-board approach that Coniglo et al. (2018) proposed for exploring whether a country's transition to the green economy followed path-dependency or not. We used the definition of new products that Coniglio et al. (2018) proposed to define new green products. In addition, the definition of RCA that is proposed by Balassa (1965) is used to define the green products in the countries' export baskets as those with RCA above 1 (Coniglio et al., 2018; Hidalgo et al., 2007). New green products can be defined as products that were not in the green production baskets of a country at time $t_0=2007$ and enter to the green production baskets of a country at time $t_1=2017$. Therefore, new green products in this research are those with a RCA lower that 0.2 at time $t_0=2007$ and above 1 at time $t_1=2017$.

Similar to Coniglio et al. (2018) and Hidalgo et al. (2007), the relatedness between any pair of products *i* and *j* that were exported in the world at time $t_0=2007$ is calculated in an MxM matrix, where the products *i* are in the rows and products *j* are in the columns of the matrix, and the values in the matrix show the relatedness between products *i* and *j*. The values of the matrix (relatedness between products *i* and *j*) are obtained as follows:

1. The RCA value for all products that is exported by each country, *c,* in the world and for time $t_0$ is calculated to "…measure whether a country, *c*, exports more of good *i*, as a share of its total exports…" (Hidalgo et al., 2007, p. 484) as follow:

$$RCA_{ci} = \frac{X_{ci}}{\sum_c X_{ci}} / \frac{\sum_i X_{ci}}{\sum_{ci} X_{ci}} \tag{1}$$

2. After calculating the RCA values for each country, *c*, in year *t*, if RCA of product *i* for country, *c*, is above 1 it means the country is a major exporter of the product *i* and it has RCA above 1 for product *i* at time *t*, otherwise the RCA is 0 (Hidalgo et al., 2007; Coniglio et al., 2018):

$$x_{ci} = \begin{Bmatrix} 1 \; if \; RCA_{ci} > 1 \\ 0 \quad otherwise \end{Bmatrix} \tag{2}$$

"where $RCA_{ci}$ is the standard Balassa (1965) index employed as a measure of export specialization." (Coniglio et al., 2018, p. 11). Then, after determining the RCA values for each country, *c*, Hausmann and Klinger's (2007) method is used to calculate the relatedness between any pair of products *i* and *j* as the minimum of the pair-wise conditional probability of being co-exported (Hidalgo et al., 2007; Coniglio et al, 2018):

$$\varphi_{i,j} = \min\{P(x_i \; / \; x_j), P(x_j \; / \; x_i)\} \tag{3}$$

where $\varphi_{i,j}$ is the relatedness values in the MxM matrix.

According to Coniglio et al. (2018), in the third step of the analysis, products that are exported with a RCA above 1 for each country, *c*, at time $t_0$ should be listed. After implementing the first two steps, we proposed $G_{ct_0}$ as the set of green products that are exported by each country, *c*, at time $t_0$=2007. Next, an MxC matrix, $D_{ic}$, of relatedness between the new green products at the time $t_0$=2007 and the initial green products of the countries at time $t_1$=2017 is developed as follows:

$$D_{ic} = \begin{Bmatrix} d_{ic}(\varphi_{i,j}) = \max(\varphi_{i,j}) \; when \; j \in G_{ct_0}, i \in N_c \\ no \; value \qquad if \; j \notin G_{ct_0} \end{Bmatrix} \tag{4}$$

where $d_{ic}(\varphi_{i,j}) = \max(\varphi_{i,j})$ shows the relatedness of new green products at time $t_1$=2017 with the most related (highest relatedness) green products at time $t_0$=2007 (Coniglio et al., 2018). The relatedness values are between 0 to 1, with the closer the relatedness value to 1, the more similar the capabilities that two products *i* and *j* require for production are.

The final step is to provide statistical analysis to explore whether the new green products that entered to the countries' export baskets followed path dependence or non-path dependence. Similar to Coniglio et al. (2018), a counterfactual distribution of relatedness for each country, *c*, was constructed by implementing the Monte Carlo simulation with "…1,000 random draws of size equal to the actual number of new…" (p. 12) green products to test Hidalgo et al.'s (2007) path-dependence hypothesis for developing countries' production basket and, at the same time, observe if there are any new green products that did not follow a path-dependent process. In essence, this compares the distribution of the new green products relatedness value that is obtained from equations (4) with the counterfactual distribution to explore the following three possible scenarios (Coniglio et al., 2018):

1. If the distribution of new green products relatedness value stands fully to the right side of the counterfactual distribution, the hypothesis of random relatedness for any level of proximities can be rejected for actual data (full path-dependence).
2. If the distribution of the new green products relatedness value stands below the counterfactual distribution, the hypothesis of random relatedness for any level of proximities cannot be rejected for actual data (no path-dependence).
3. If the distribution of the new green products relatedness value stands partially to the right side of the counterfactual distribution, the hypothesis of random relatedness can be rejected for new green products with distribution stands above counterfactual distribution (path-dependence and non-path dependence process).

In accordance with Duranton and Overman's (2005) and Coniglio et al.'s (2018) methods, this analysis is implemented using kernel smoothed density estimation as:

$$\bar{K}(d) = \frac{1}{(\sum_{i=1}^{M} \sum_{i=2007}^{2017} I_{it})h} \sum_{i=1}^{M} \sum_{i=2007}^{2017} I_{it} f\left(\frac{d-d_{it}}{h}\right) \text{ for all countries, c} \qquad (5)$$

where "…densities [are] calculated non-parametrically using a Gaussian Kernel function with bandwidth h set according to Silverman's optimal rule of thumb." (Coniglio et al., 2018, p. 13). In equation (5), $d_{it}$ is obtained using equation (4), and $\sum_{i=1}^{M} \sum_{i=2007}^{2017} I_{it}$ equals to the total number of green products. In addition, a regression model is developed to explore if the green growth indicators (development of environmental-related technologies per capita, number of patents related to environmental-related technologies, and development of environment-related technologies, as a % of all technologies) obtained from OECD statistics database have a significant effect on a country's green production expansion. This model is discussed in depth in the "High investments structural jump" section.

**Results**

Figure 1 shows the kernel distribution of relatedness, as defined in equation 4, between new green products at time $t_1$=2017 and the products with RCA>1 at $t_0$=2007 for all countries.

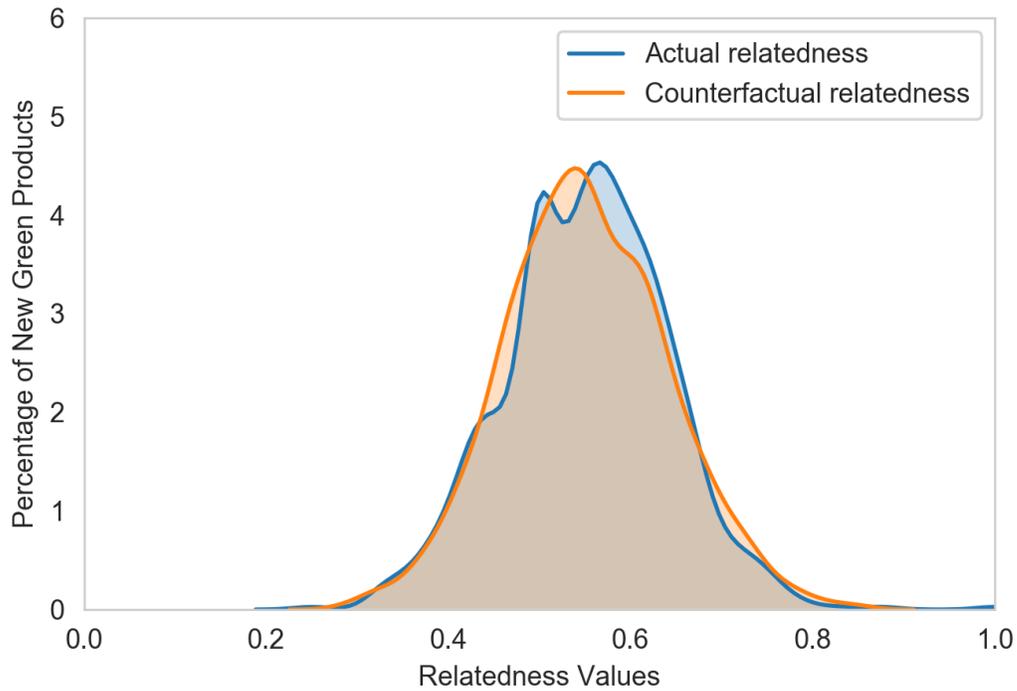

**Figure 1. Kernel distribution of relatedness between new green products at time $t_1=2017$ and time $t_0=2007$: actual new green products data, versus random data**

The horizontal axis in Figure 1 shows the relatedness values – 0 to 1- between new green products at $t_1=2017$ and the products with RCA>1 at $t_0=2007$ for all countries. The blue kernel distribution shows the relatedness between new green products at $t_0=2007$ and the products with RCA>1 at $t_0=2007$, while the orange kernel distribution shows the randomly generated relatedness values that were obtained as a result of the counterfactual analysis.

The comparison between the blue kernel distribution and orange kernel distribution shows whether the countries green growth followed path-dependence process or not. The comparisons show that countries' diversification to a green economy followed the path-dependence process when the relatedness values between 0.58 to 0.7. The relatedness values above 0.7 demonstrate that the non-path dependent process is followed when the products have high degree of relatedness.

This shows that countries did not enhance their green production baskets based on the products for which they already had an RCA>1 and they jumped in their PS network.

The comparison also shows that countries' diversification to green growth followed the non-path dependent process for a considerable number of green products (represented by the orange area above the blue area). Thus, our results show that countries did jump in their PS and produced green products that did not share similar capabilities with their existing green production baskets.

According to Mealy and Teytelboym (2018), China ranked first in terms of green growth, and China followed Hidalgo et al.'s (2007) concept of path–dependence to develop their green production baskets.

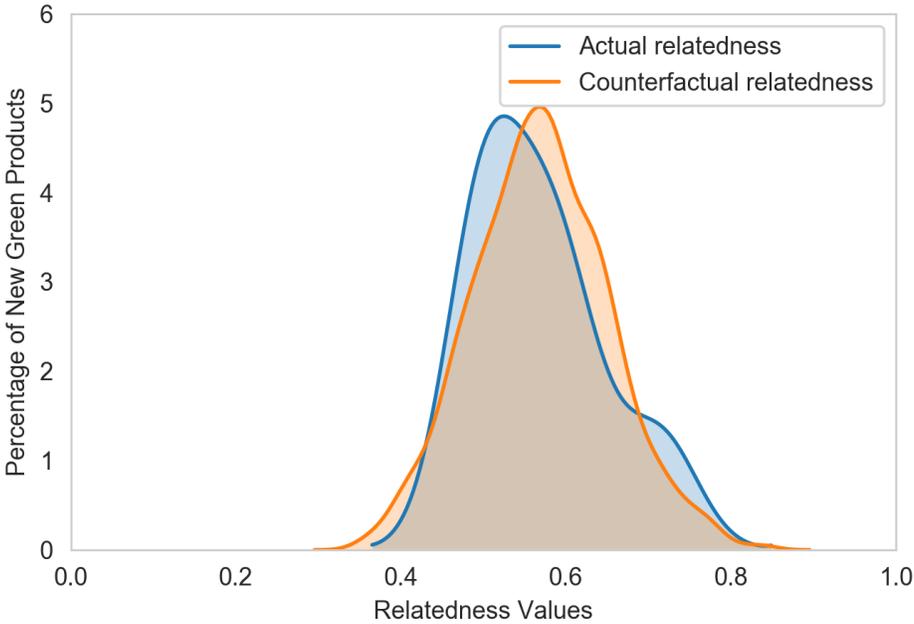

**Figure 2. China's kernel distribution of relatedness between new green products at time $t_1$=2017 and time $t_0$=2007: actual new green products data versus random data**

Similar to Figure 1, the horizontal axis in Figure 2 shows the relatedness values – 0 to 1- between new green products at $t_1$=2017 and the products with RCA>1 at $t_0$=2007 for China. By comparing the kernel distribution of relatedness between new green products at $t_1$=2017 and the products with RCA>1 at $t_0$=2007, the blue kernel distribution, with the kernel distribution of randomly generated relatedness for China's green product relatedness values, the orange kernel distribution, it is shown that China followed the non-path dependence process (the orange area stands above blue area) for some green products, but at the same time, followed the path-dependence process (the blue area that stands partially to the right side of orange area), especially for new green products that the relatedness values above 0.7, which demonstrate that the path-dependency process is followed when the products have high degree of relatedness, aligned with Hidalgo et al.'s (2007) hypothesis of path-dependence. This shows that China enhanced its green production baskets based on the products for which it already had an RCA>1. That is, China exhibits patterns that both adhere to and deviate from the path dependency hypothesis. This shows China's structural change in their production capabilities enabled them to produce more green products and grow their economy while reducing their environmental risks (Haibo et al., 2019; Li et al., 2019; Pan et al., 2019).

**High-investments structural jumps**

In the previous section it was found that countries not only follow the path-dependent process to expand their green production baskets, but also follow the non-path dependent process to produce a considerable amount of green products. A country's ability to expand its product basket without being limited by its current capabilities is demonstrated by its ability to expand to sectors of production unrelated to its current production basket; countries capable of this are better equipped for structural transformations and rapid economic growth (Coniglio et al., 2018).

The goal of this section is to identify factors which enable jumps in the green production basket. To this end, a linear regression model is used where the dependent variable is the share of the new green products added to countries export basket following the path-dependent process. As defined in the methods section, new green products are green products with RCA<0.5 in 2007 and above 1 in 2017. The linear regression model is

$$Y = \alpha + \beta X_{i,t_0} + \varepsilon_{it}$$

where $X_{i,t_0}$ is the independent variable that is calculated for each country at time $t_0$=2007, and $\varepsilon_{it}$ is the error term. The dependent variable and independent variables are described in Table 2.

Table 2. Description of dependent and independent variables

| Variables | Description | Source |
|---|---|---|
| Dependent variable: New green products | Share of new green products (RCA<0.5 at $t_0$=2007 and RCA>1 at $t_1$=2017) over the total number of green products | Calculations described within this paper |
| Independent variable: Development of environment-related technologies per capita | Number of environmental-related inventions per million residents of each country | Organization for Economic Co-operation and Development (OECD) |
| Independent variable: Number of patents related to environmental-related technologies | Number of patents that each country reported related to environmental-related technologies | Organization for Economic Co-operation and Development (OECD) |
| Independent variable: Development of environment-related technologies, % all technologies | Number of environmental-related technologies that countries developed compared to all developed technologies | Organization for Economic Co-operation and Development (OECD) |

It is expected to see that the independent variables have a negative effect on the dependent variable, that is, as countries introduce more patents and develop more environmental-related technologies, their green production basket diversifies regardless of their existing production capabilities. The result of the regression analysis is provided in Table 3.

Table 3. The factors that cause countries jump in their green production basket

|  | (1) | (2) | (3) |
|---|---|---|---|
| Development of environment-related technologies per capita | -0.00012** | -1.080* | -1.108* |
| Number of patents related to environmental-related technologies |  | -1.826 | -1.705 |
| Development of environment-related technologies, % all technologies |  |  | -1.409 |
| Number of countries | 65 | 65 | 65 |

*Note:* $^*p<0.05$, $^{**}p<0001$,

As expected, all three factors have a negative effect on producing the new green products following the path-dependent process. This shows that as the values of the independent variables increase, the less dependencies exist for a country's green production. As countries invested more in the development of environmental related technologies, they jumped in their production basket and were able to produce more green products. This is the main reason that China had a jump in its green production basket and produced more green products.

**Discussion**

The results of this research showed that the 65 countries grew their green economy by following both path dependence and non-path dependence processes. We used China as our case study since it had the largest green growth compared to other countries in the world (Mealy and Teytelboym, 2018), and found that China did jump—that is, followed non-path dependence process for considerable number of their green products—in its PS, and could produce more green products compared to all other countries. This shows that China implemented successful structural reforms in order to grow while reducing its environmental risks, especially climate change (Haibo et al., 2019; Li et al., 2019; Pan et al., 2019; Shen et al., 2018; Li et al., 2018). Additionally, the results showed that innovating and developing new environmental-related technologies has a significant effect on the new green products that countries produce: they become less dependent on their existing capabilities. This confirms that as countries innovate and develop more new environmental-related technologies, they can jump in their product space and produce new green products that are not related to their existing production capabilities.

The results of this paper show that even as "…governments around the globe are more and more seduced by the PS idea of 'latent comparative advantage', which suggests that policy effort should be 'smartly' targeted to those products that are not yet in countries' export baskets but are related to it (i.e. small jumps over the PS are those that are likely to be effective)" (Coniglio et al., 2018, p. 28), countries diversify their green production basket also based on large PS jumps as a result of the non-path dependent process. This shows that countries can plan to jump in their PS to produce more green products by having a "…better endowment of human…" (Coniglio et al., 2017, p. 28) and "[natural] capital to develop their comparative advantage in new areas of the PS" (Coniglio et al., 2018, p. 28), and better intervention from governments (Mealy and Teytelboym, 2018) in order to support industries, and thus make such high-investment structural jumps in green production.

**Conclusion**

An economy grows by upgrading the products they produce and export (Hidalgo et al., 2007). Countries follow the path-dependence (Hidalgo et al., 2007) or non-path dependence process (Coniglio et al., 2018) to grow their economy. That is, if a country uses its existing capabilities to grow its economy, it follows the path-dependence process, and if a country produces a product that uses divergent capabilities from its existing production basket, it follows the non-path dependence process. Following the non-path dependence process requires the country to make structural changes in its production basket (Coniglio et al., 2018). A country's economic growth will increase its environmental risks, specifically climate change, therefore, it is crucial for environmental risks to be minimized even as a country seeks to grow (Knight and Schor, 2014).

Economic growth and economic development causes countries to face several problems including the aforementioned environmental risks, and also high energy and natural resource consumption. Climate change is the most considerable problem that emerges when countries try to grow their economy. However, a transition to a green economy enables countries to grow their economy while reducing this environmental risk. The transition to a green economy requires structural changes in industry, and the implementation of policy for is a challenge for all countries, particularly poor countries. The results of this research showed that China, the country that has the largest growth in its green economy between 2007 and 2017, jumped in its green production basket and successfully completed major structural changes in its industry. In addition, the results of this research provide a better understanding on how countries enhanced their green growth, and thus, how other countries could accomplish similar actions to create or improve their green economy development plan while reducing their environmental risks, especially climate change. We characterize green growth as exhibiting dual dynamics: (1) path-dependent growth that "exploits" current infrastructure, and (2) high-investment structural growth that "explores" new structural changes via strategic investments. Further, development of these framework will allow countries or regions to strategically promote "path-dependent growth" or "high-investment structural growth" in order to achieve their green economy goals. In addition to green economy, the method used in this paper can be used for different areas, such as understanding the pattern of industrial evolution in a region or country, or understanding the pattern of technological relatedness evolution in a region or country.